# Charge Transfer Kinetics at the Solid-Solid Interface in Porous Electrodes

Peng Bai[1] and Martin Z. Bazant*,[1,2]

[1]Department of Chemical Engineering and [2]Department of Mathematics, Massachusetts Institute of Technology, 77 Massachusetts Avenue, Cambridge, Massachusetts 02139, USA

* Corresponding author: bazant@mit.edu

**ABSTRACT**. Interfacial charge transfer is widely assumed to obey Butler-Volmer kinetics. For certain liquid-solid interfaces, Marcus-Hush-Chidsey theory is more accurate and predictive, but it has not been applied to porous electrodes. Here we report a simple method to extract the charge transfer rates in carbon-coated $LiFePO_4$ porous electrodes from chronoamperometry experiments, obtaining curved Tafel plots that contradict the Butler-Volmer equation but fit the Marcus-Hush-Chidsey prediction over a range of temperatures. The fitted reorganization energy matches the Born solvation energy for electron transfer from carbon to the iron redox site. The kinetics are thus limited by electron transfer at the solid-solid (carbon-$Li_xFePO_4$) interface, rather than by ion transfer at the liquid-solid interface, as previously assumed. The proposed experimental method generalizes Chidsey's method for phase-transforming particles and porous electrodes, and the results show the need to incorporate the Marcus kinetics in modeling batteries and other electrochemical systems.

## Introduction

Electrochemical energy systems are key enabling technologies for renewable energy, electrified transportation, and smart grids, in which energy conversion and delivery are carried out by Faradaic reactions between electrons and ions.[1] In many systems, such as lithium ion batteries and solid oxide fuel cells, charge transfer occurs at complex solid-solid interfaces in porous electrodes, coupled to non-equilibrium thermodynamics.[2] This greatly complicates the interpretation of electrochemical measurements, compared to the simple case of uniform, flat liquid-solid interfaces,[3, 4] widely studied in electroanalytical chemistry.

The importance of interfacial charge transfer kinetics is illustrated by lithium iron phosphate ($LiFePO_4$),[5] one of the most intensively studied cathode materials for Li-ion batteries. Its electrochemical performance can be dramatically improved by various surface modification techniques, such as carbon coating[6] and graphene wrapping[7] for increasing the electronic conductivity, and anion adsorption[8] for reducing the ionic energy barrier. The basic idea behind these surface treatments is that interfacial charge transfer involves both lithium ions and electrons.

As shown in Fig. 1, when a negative overpotential is applied to a carbon-coated $Li_xFePO_4$ crystal, lithium ions jump into the interstitial vacancies in the first atomic layer of the crystal while the electrons tunnel to the iron site and reduce $Fe^{3+}$ to $Fe^{2+}$. After the charge-transfer reaction, adjacent $Fe^{2+}$ and $Li^+$ ions with the local distortion around them form a neutral quasiparticle, or polaron, that can diffuse into the crystal.[9] Since Li$_x$FePO$_4$ is a poor electronic conductor,[10, 11] the carbon coating acts as the "electrode" that provides electrons for the electrochemical reaction. On the other hand, carbon as an anode material barely accommodates lithium ions at voltages larger than 2V,[12] so carbon layers coated on cathode materials working at 3-4V provide a purely ionic barrier against lithium transfer between the solid and the electrolyte, which prevents lithium intercalation across thick carbon coatings.[10, 11]

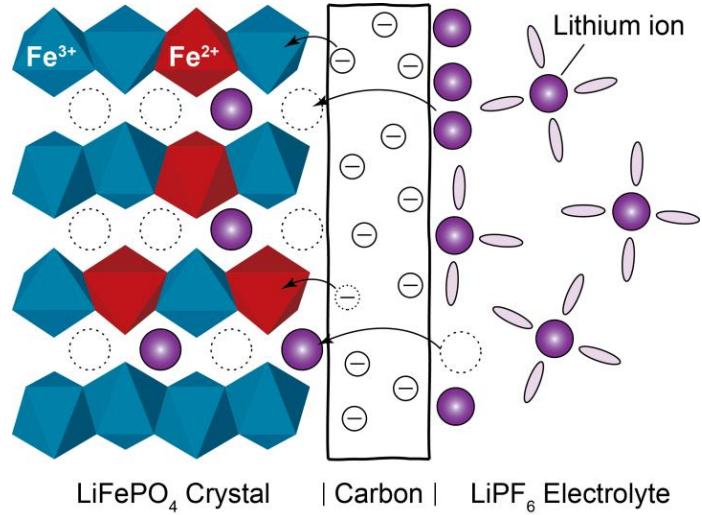

**Figure 1. Schematic demonstration of the interface of a carbon-coated Li$_x$FePO$_4$ crystal.** Lithium ions in the electrolyte jump across the carbon coating into the vacancies in the first atomic layer of the crystal, while electrons in the carbon coating tunnel to the adjacent iron site to reduce the Fe$^{3+}$ ions. (PO$_4$ tetrahedrons are omitted for clarity.)

Existing mathematical models neglect the details of charge transfer and assume that the net reaction, Li$^+$ + e$^-$ + FePO$_4$ ↔ LiFePO$_4$, obeys phenomenological Butler-Volmer (BV) kinetics,[1] focusing on the role of the lithium ion.[13, 14] The net reaction rate $k$ for single charge transfer is expressed as,

$$k(\eta)/k_0 = \exp[-\alpha\eta] - \exp[(1-\alpha)\eta] \qquad (1)$$

where $k_0$ is the exchange rate constant and $\eta = e(E - E^{0'})/k_BT$ is the dimensionless overpotential, scaled to the thermal voltage $k_BT/e$, (where $e$ is the elementary charge, $k_B$ Boltzmann's constant, and $T$ the absolute temperature). The overpotential is defined as the difference between the electrode potential $E$ and the formal potential $E^{0'}$.[15] The charge transfer coefficient, $\alpha$, is usually set to 0.5 in battery modeling. For BV kinetics, the Tafel plot of ln$k$ versus $\eta$ is a straight line of slope $-\alpha$ for $\eta<0$ and $1-\alpha$ for $\eta>0$. In principle, the fundamental rate constant $k_0$ can be determined from the y-intercept of the fitted Tafel line, but the vast range of fitted exchange currents for the same material[13, 16-19] (over seven orders of magnitude, 10$^{-6}$ to 10$^1$ A m$^{-2}$) undermines the validity of this approach.

The fact that increasing the electronic conductivity of LiFePO$_4$ (an insulator) can dramatically improve the high-rate performance suggests that lithium intercalation reaction may instead be limited by electron transfer between the carbon coating and the redox site in the crystal,

$$\text{Fe}^{3+} + e^- \leftrightarrow \text{Fe}^{2+} \qquad (2)$$

while the ion transfer reaction, $Li^+ \leftrightarrow Li^+_{ads}$, is fast. In this case, Marcus theory,[3, 20, 21] honored by the Nobel Prize in Chemistry,[3] could be applied to fit curved Tafel plots for a solid-solid interface (to our knowledge, for the first time), if the fundamental rate constants could be unambiguously extracted from porous electrode measurements.

In this work, we propose a statistical method of Tafel analysis that can be used to extract fundamental reaction rates for porous electrodes. Since the material properties of $LiFePO_4$ are readily available, we use typical commercial $LiFePO_4$ to validate our method. The experimental results are in excellent agreement with microscopic electron transfer theory, thus shedding new light on Faradaic reactions at solid-solid interfaces.

## Results

**Curved Tafel plots of total currents.** As motivation, we begin by performing classical Tafel analyses of literature data. In a recent experiment, Munakata et al.[22] tested a single 20μm agglomerate of carbon-coated $LiFePO_4$ over a range of constant currents, up to 900 C (4-second discharge). The measured voltage drops at half fillings of the porous particle under different currents yield a highly curved Tafel plot, where the $\alpha$=0.5 BV slope can only fit a small portion of the data.[22] Since the voltage drops at high currents are easily affected by concentration polarization (transport limitation), we read the voltage drops at the lowest fillings of the discharge plateaus, and plot the data in Fig. 2a, obtaining another curved Tafel plot. Analysis of the discharge curves from Kang and Ceder[23] shows a similar trend in Fig. 2b. This as-yet unexplained phenomenon is not unique to $LiFePO_4$, but has also been reported for $Li-O_2$ batteries by extrapolating initial voltage drops under different currents.[24]

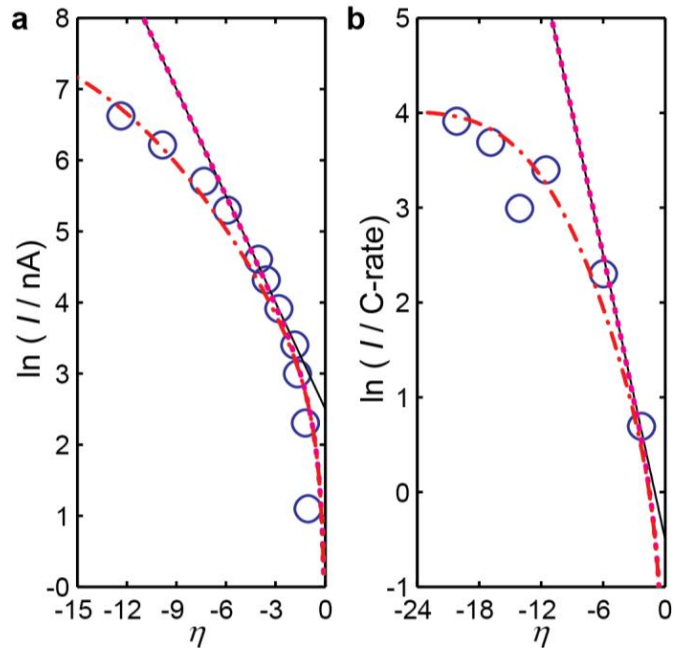

**Figure 2. Tafel analyses of discharge currents against corresponding voltage drops.** (a) Data of a 20-micron $LiFePO_4$ secondary particle from Munakata et al.[22] (b) Data of a coin cell battery with nano-$LiFePO_4$ particles from Kang and Ceder.[23] Voltage points are taken at the beginnings of the discharge plateaus. Both data deviate from the Butler-Volmer model (dotted line) and the Tafel slope (thin solid line) with $\alpha$=0.5, which however can be fitted by the Marcus-Hush-Chidsey model $k_{red}(\lambda,\eta)-k_{ox}(\lambda,\eta)$ with $\lambda^{(a)} = 15$ and $\lambda^{(b)} = 12$ (dot-dashed curves).

The total currents used in standard Tafel analyses,[1, 22] however, do not accurately represent the fundamental reaction rates since the active area of a porous electrode is non-uniform and varies with the applied current.[25] Moreover, the method of estimating the activation overpotential from the overshoot of the voltage plateau is fundamentally flawed, since the plateau is an emergent property of collection of phase-transforming particles.[26] Of course, this method also cannot be applied to solid-solution materials without a voltage plateau.

**Transient currents and reaction rates.** Curved Tafel plots have been reported in many surface bound redox systems since the seminal work by Chidsey,[4] who extracted reaction rates from voltage-step chronoamperometry experiments by fitting the linear relationship shown in the semi-logarithmic plot of the transient current versus time.[4] The transient current after a voltage step is fitted to a simple exponential decay,[27]

$$I = e\Gamma k_{app} \exp(-k_{app} t) \quad (3)$$

where $\Gamma$ is the coverage of the electrode and $k_{app}$ is the single decay rate, assumed to represent the rate of reactant consumption. Chidsey's method ensures that the kinetics under different potentials are clearly separated and thus avoids the ambiguity of picking voltage points from constant current discharges.

Although Eq. (3) can be justified for flat electrodes with uniform reaction rates, such as the surface-bound monolayers of redox species investigated by Chidsey, it cannot be applied to porous electrodes or phase transforming particles with non-uniform reaction rates. Mosaic instability throughout the electrode[25, 26] results time-dependent populations of reacting particles, or equivalently an evolving active internal area. In other words, $\Gamma$ in Eq. (3) is not constant for electrodes with complex thermodynamics.

Recently, Bai and Tian[28] proposed a simple statistical model to describe discrete-particle phase transformations in porous electrodes. Assuming a simple three-state Markov chain of untransformed, transforming, and transformed particles, the total current of a porous electrode is proportional to the population of phase-transforming particles.[28] The transient current in response to a voltage step can be expressed as,

$$I = kQ \left[ \frac{N_0 k + (N_1 - 1)k_A}{k - k_A} \exp(-kt) + \frac{(1 - N_0 - N_1)k_A}{k - k_A} \exp(-k_A t) \right] \quad (4)$$

where $Q$ is the capacity of the electrode; $k$ is the reaction rate at the surface of phase-transforming particles that continuously accept (release) lithium ions during discharge (charge); $k_A$ is a generalized activation rate, which is identical to the nucleation rate at low overpotentials for phase-separating materials, but also captures the random activation process at high overpotentials when phase separation is suppressed.[14, 29] ($k$ and $k_A$ are identical to $\bar{m}$ and $\bar{n}$ in ref. 28, respectively.) $N_0$ is the initial fraction of transforming particles (reacting areas) in the electrode. $N_1$ is the initial fraction of transformed particles (inactive areas) in the electrode, which is set to zero in the present work.

Since the statistical model does not account for the microscopic dynamics of single-particle transformations, the generalized activation rate $k_A$ simply reflects the population of reacting particles or sites, or more generally, the evolving active surface area of the porous electrode. Although this assumption may seem overly simple, the statistical model was recently validated by Levi et al.[30] by fitting *in situ* electrochemical quartz admittance data "surprisingly" well for a thin microarray of LiFePO$_4$ particles. Interestingly, in the limit of fast activation ($k_A \to \infty$), the full internal surface of the porous electrode becomes active, or equivalently, in the monolayer limit,[4] equation (4) reduces to $I=kQ\exp(-kt)$, identical to equation (3) for surface bound redox systems.[27] For our porous electrodes, fitting $k_A$ and $N_0$ enables more accurate determination of the reaction rate $k$.

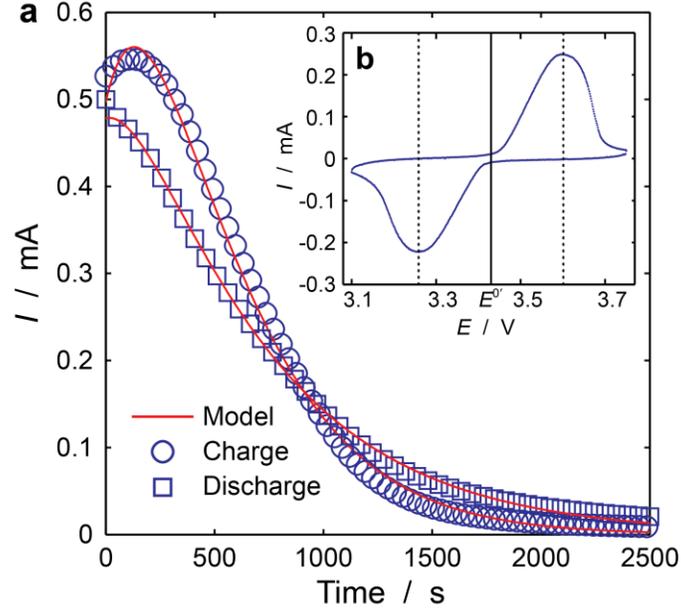

**Figure 3. Typical experimental results of the LiFePO$_4$ coin cells.** (a) Examples of the transient currents with fitting curves. The parameters for the charging step (181mV) are $k$=0.003088 s$^{-1}$, $k_A$=0.00325 s$^{-1}$, $Q$=0.4245 As and $N_0$=0.3789, where the measured capacity is 0.4292 As. Parameters for the discharging step (−196 mV) are $k$=0.001598 s$^{-1}$, $k_A$=0.00515 s$^{-1}$, $Q$=0.4006 As and $N_0$=0.747, where the measured capacity is 0.4449 As. The size of the voltage step was calculated with respect to the formal potential $E^{0'}$=3.430V, defined as the average of the two peak potentials in the (b) Cyclic voltammetry of our LiFePO$_4$ coin cells scanned at the rate of 0.1mVs$^{-1}$.

Figure 3a provides typical transient currents of a LiFePO$_4$ coin cell used in our experiments (see Methods for details). By fitting the transient currents with equation (4), we can extract the reaction rates $k$ for different overpotentials. Although the value of the total capacity $Q$ can be estimated from experiments, it is relaxed in the fitting to account fluctuations in experimental conditions and errors in time integration of the transient current, illustrated in Fig. 3.

**Curved Tafel plot of reaction rates.** Fundamental reaction rates extracted from transient currents in many different voltage-step experiments are shown in Fig. 4a. The data clearly deviates from the linear Tafel dependence of the BV equation at overpotentials larger than $4k_BT/e \approx 100$mV. As motivated above, we adopt instead the Marcus-Hush-Chidsey (MHC) model[4, 31, 32] to calculate the theoretical reaction rates of equation (2),

$$k_{\text{red/ox}}(\lambda,\eta) = A\int_{-\infty}^{+\infty} \exp\left\{-\frac{(x-\lambda\pm\eta)^2}{4\lambda}\right\}\frac{dx}{1+\exp\{x\}} \qquad (5)$$

where $\lambda$ is the dimensionless reorganization energy scaled to $k_BT$ and $\eta = e(E-E^{0'})/k_BT$ is the same dimensionless overpotential in equation (1). The integral over the dimensionless variable, $x=(\varepsilon_{el}-eE)/k_BT$, accounts for the Fermi statistics of electron energies, distributed around the electrode potential.[4] $A$ is the pre-exponential factor, accounting for the electronic coupling strength and the electronic density of states (DOS) of the electrode. For non-adiabatic electron transfer at a carbon electrode,[33] the DOS varies with the electrode potential[33] and cannot be placed outside the integral, but within the small potential range considered in this work, the DOS of the carbon coating is roughly constant[33] as assumed in equation (5). The net reaction rate of lithium insertion is $k(\lambda,\eta)=k_{red}(\lambda,\eta)-k_{ox}(\lambda,\eta)$, and the corresponding rate constant can be calculated as $k_0=Ak_{red/ox}(\lambda,0)$.

The extracted reaction rates are compared to various theoretical models. Curves calculated from the MHC model (equation (5)) are plotted in Fig. 4 and adjusted to ensure the convergence to the BV curves at low overpotentials. Remarkable agreement is observed between the MHC curves for $\lambda=8.3$ and the reaction rates for both the charge and discharge processes, across the entire range of overpotentials. Curves calculated from the classical Marcus rate equation (without integrating against the Fermi distribution) are also plotted, but the agreement is not as good, as expected for electron transfer from a metal electrode. Since the Marcus rate reaches a maximum and then decreases for $|\eta|>\lambda$ (the celebrated "inverted region", observed in bulk charge-transfer reactions), a rather large reorganization energy ($\lambda=13.5$) is needed to fit the data.

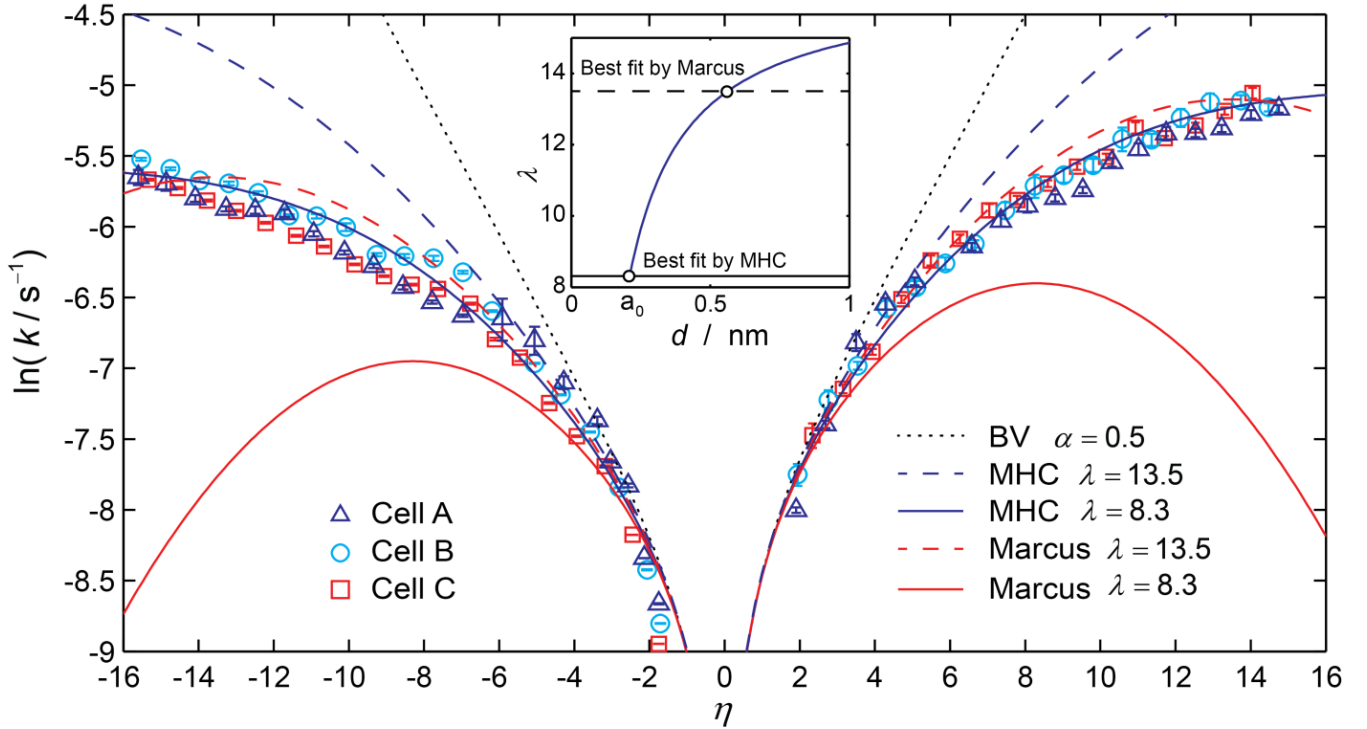

**Figure 4. Tafel analysis of the reaction rates (symbols) extracted by equation (4) from chronoamperometry experiments of three coin cells at room temperature.** Error bars indicate the 95% confidence bounds. Curves are theoretical results from the BV model (dotted black), the MHC model (blue), and the classic Marcus model (red). Overpotentials are scaled to the thermal voltage $k_BT/e\approx25.7$mV. Inset: dependence of the reorganization energy on the separation distance between the reactant and electrode surface, i.e. equation (6).

Besides capturing the over-potential dependence of the reaction rates, microscopic theories of charge transfer are able to predict the reorganization energy from first principles, without any empirical fitting. For a liquid-solid interface, the total reorganization energy for electron transfer has two main contributions:[15] the "outer" reorganization of the solvent, $\lambda_o$, dominated by long-range electrostatic forces, and the "inner" relaxation of the reactant itself, $\lambda_i$, dominated by short-range bond forces. Many experiments have shown that the former dominates, $\lambda \approx \lambda_o$.[27] This approximation should also hold, if not better, for reactions at a solid-solid interface, since the "reactant" is part of the same solid dielectric continuum assumed for the "solvent", and the changes of the bonds connecting the "reactant" and the "solvent" are highly cooperative.[34] We thus approximate the (dimensionless) reorganization energy by the Born energy of solvation,[15, 20]

$$\lambda \approx \lambda_o = \frac{e^2}{8\pi\varepsilon_0 k_B T}\left(\frac{1}{a_0} - \frac{1}{2d}\right)\left(\frac{1}{\varepsilon_{op}} - \frac{1}{\varepsilon_s}\right) \tag{6}$$

where $\varepsilon_0$ is the permittivity of free space, $a_0$ the effective radius of the reactant, $d$ the distance from the center of the reactant to the surface of the electrode, $\varepsilon_{op}$ the optical dielectric constant, and $\varepsilon_s$ the static dielectric constant. We use the length of Fe-O bond in the $FeO_6$ octahedron as the radius $a_0$, which is roughly 0.21nm.[35] The dielectric constants are available from first-principles calculations:[35] $\varepsilon_{op} \approx \varepsilon_{inf} = 4.74$ and $\varepsilon_s = 11.58$. If we assume that the $FeO_6$ octahedrons were in direct contact with the carbon coating, i.e. $d=a_0=0.21$nm, then the calculated reorganization energy is 213 meV, corresponding to (dimensionless) $\lambda=8.3$ at room temperature. If the $FeO_6$ octahedrons were separated from the carbon coating by $PO_4$ tetrahedrons, adding the length of the O-O bond (~0.23nm), i.e. $d=0.44$nm, gives us $\lambda=12.64$. Surprisingly, the MHC model with the lower bound $\lambda=8.3$ can accurately fit the reaction rates extracted from three coin cells, which is a striking validation of the theory.

**Table 1. Fitted rate constant $k_0$ for curves in Fig. 4**

| Model | $\lambda$ | $k_0$ ($10^{-4} s^{-1}$) | |
|---|---|---|---|
| | | $\eta<0$ | $\eta>0$ |
| BV | $\alpha=0.5$ | 1.174 | 2.035 |
| MHC | 8.3 | 1.190 | 2.062 |
| | 13.5 | 1.208 | 2.093 |
| Marcus | 8.3 | 1.204 | 2.086 |
| | 13.5 | 1.204 | 2.086 |

As listed in Table 1, the fitted rate constants for discharge (cathodic) are slightly smaller than those for charge (anodic). The broken symmetry may be attributed to the different lithium surface compositions of the active surface for adsorption and desorption,[36] and could also be explained by nonlinear concentration and stress effects on the exchange current in generalized Marcus kinetics for mixed ion-electron transfer processes in solids.[2] Because the reaction rates are inferred from the decay rates of the transient currents via equation (4), they are independent of the absolute magnitudes of the currents, which depend on the population of transforming particles. As such, the $k_0$ values in Table 1 can be viewed as effective material properties of the active particles, independent of the areas of reacting surfaces in the porous electrodes, although they do reflect some averaging over different local surface concentrations during a single-particle phase transformation. Since the area of each lithium site is ~0.5× 0.5 nm$^2$, $k_0$=2.0×10$^{-4}$ s$^{-1}$ (per site) yields an exchange current density $I_0$≈1×10$^{-4}$ A m$^{-2}$, if each particle transforms homogeneously, as expected for high rates.[14,29]

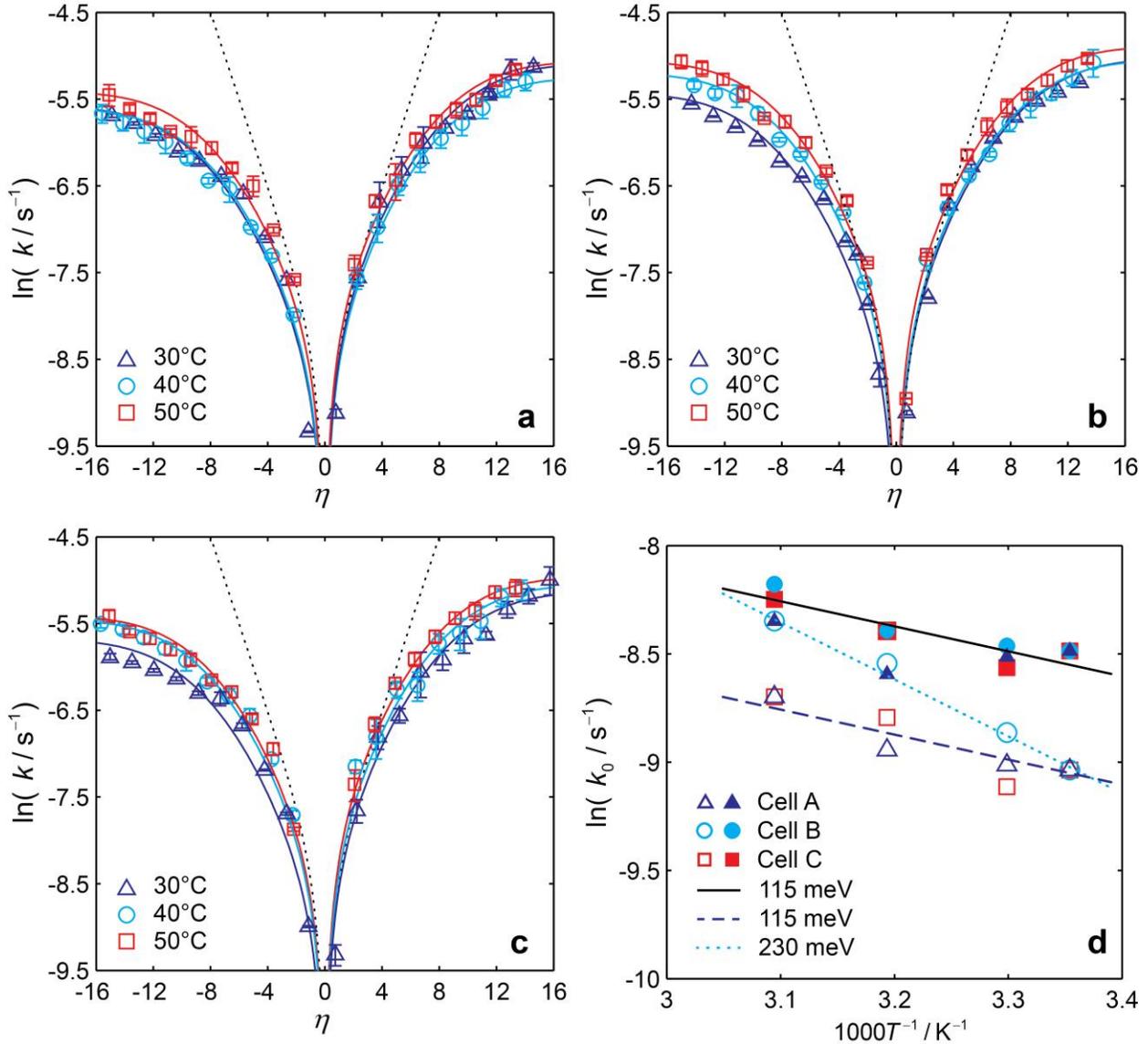

**Figure 5. Temperature dependence of the fundamental reaction rates.** Tafel plots of (a) cell A, (b) cell B and (c) cell C at 30°C, 40°C and 50°C. Dotted curves indicate the BV model (linear Tafel plot) with $α$=0.5. Solid curves result from fitting MHC model with a unique value of the reorganization energy. (d) Arrhenius plot of the rate constants under 25°C, 30°C, 40°C and 50°C. Open and filled symbols denote cathodic (discharge) and anodic processes, respectively.

The reaction rates extracted from experiments at higher temperatures are displayed in Fig. 5 for each coin cell. In a striking validation of the MHC theory, the nine series of reaction rates can all be fitted by the same reorganization energy (214±1 meV), independent of temperature, with 0.5% accuracy. Further validation comes from Arrhenius temperature dependence of the corresponding rate constants ($k_0$), which yields the effective energy barriers for interfacial charge transfer reaction at zero overpotential. According to the classical Marcus theory, the effective energy barrier is roughly equal to $\lambda/4$, so the value $\lambda=13.5$ fitted in Fig. 4 yields a barrier of ~87 meV. Despite that the effective energy barriers from the secondary fitting ($k_0$ vs. $T$) could easily drift away from the true value, the barriers obtained (115 meV) from the Arrhenius plot (Fig. 5d) are consistent with this estimation. The effective barrier (230 meV) fitted from the cathodic (discharge) rate constants of cell B is comparable to the first-principles energy barrier for polaron transfer in bulk crystal, which under the "adiabatic" assumption[9] yields a diffusion coefficient of order $10^{-8}$ cm$^2$s$^{-1}$. The diffusion time through micron-sized particles at such a high diffusivity is of order 1 second, whereas the time constants observed in our experiments are of order $10^4$ s. Therefore, fast adiabatic polaron transfer within the solid crystal cannot be the rate-limiting step, as further explained in the Discussion below. Instead, the small rate constants indicate very weak electronic coupling[15, 20, 33, 37] between the carbon coating and the LiFePO$_4$ crystal, consistent with non-adiabatic electron transfer kinetics at the interface. Our experiments also reveal that the reaction rate is higher in the fresh cells than in the aged cells, which may be ascribed to interfacial changes due to side reactions and elastic strain fatigue.

**Discussion**

Ionic transport may affect the overpotential via logarithmic concentration terms in the Nernst voltage,[2, 15, 37] but this cannot be the rate-limiting step during battery charge or discharge in our experiments. Taking discharge as an example, the whole electrochemical process consists of three steps: lithium ion transport in the electrolyte, charge transfer reaction at the interface, and polaron diffusion in the crystal. The characteristic time constant for lithium ions in the electrolyte to transport across a 5μm thick porous electrode is of order 0.1 second at the diffusivity ($10^{-6}$ cm$^2$ s$^{-1}$) obtained from experiments.[38] The time constant for polarons in the crystal to diffuse through a 1μm long [010] channel is strongly temperature dependent and of order 1 second at room temperature, using the diffusivity ($10^{-8}$ cm$^2$ s$^{-1}$) from first principles calculations.[35, 39-42] The capacity of our coin cells at 0.1C is ~150mAh g$^{-1}$, which is roughly 88.24% of the theoretical capacity (170mAh g$^{-1}$). According to the analysis of Malik et al.[41] (see equation (2) and figure 2 in ref. 41), such a high capacity indicates less than 0.1% channel-clogging defects in our 1μm sized particles, so that the diffusivity should stay close to the theoretical value. In summary, both transport steps appear too fast to be responsible to the decaying rates smaller than $10^{-2}$ s$^{-1}$, or equivalently time constants larger than 100s. This leaves the charge transfer reaction as the only possibility for the rate-limiting step, which also explains the negligible temperature dependence in the curvature of the Tafel plots.

Since the time constant for charge transfer reactions under small overpotentials are four orders larger than that of the solid diffusion ($10^4$ s vs. 1 s), diffusivities measured from electroanalytical methods[43-45] ($10^{-17}$-$10^{-12}$ cm$^2$ s$^{-1}$) are likely affected by the slow surface reaction, and therefore, are orders smaller than the theoretical predictions ($10^{-8}$ cm$^2$ s$^{-1}$). The same is true of diffusivities obtained by fitting classical porous electrode models[13,16] ($10^{-13}$ cm$^2$ s$^{-1}$), which must find solid diffusion limitation due to the flawed assumption of fast Butler-Volmer kinetics at high rates. We are also aware of the small theoretical diffusivities ($10^{-17}$ cm$^2$ s$^{-1}$ and $10^{-12}$ cm$^2$ s$^{-1}$) for the adsorbed lithium ions hopping between channel entrances on the surface of the FePO$_4$ crystal.[42] However, while lithium ions can find alternative path in the electrolyte to quickly reach the favorable entrance, there is no quicker path for electrons to jump between the solid crystal and the carbon coating. Surface modifications may lower the energy barrier for surface diffusion,[8] but will also inevitably alter the reorganization energy for the electron transfer reaction. Our method may be applied to further distinguish the difference.

Besides the fast diffusion of solid-state lithium ions (polarons) along the [010] channels,[35, 39-41] the de-wetting of lithium ions from the largest and most active (010) facet of $FePO_4$ particles to lower the surface energy[36, 46] implies that constant low concentration of intercalated lithium is maintained at the interface. We conclude that small variations of the concentrations in the logarithm term should not cause significant voltage change,[2, 27] consistent with the success of our reaction-limited model. Voltage fluctuations due to the contact resistance are also negligible. The contact resistances of the coin cells measured by potentiostatic impedance are around 6Ω, which would at most cause a 9mV voltage drop at a step of 392mV, or a 0.6mV voltage drop at a step of 74mV.

The excellent agreement between the data under various temperatures and the widely used Marcus-Hush-Chidsey model justifies *a posteriori* our method of extracting reaction rates from chronoamperometry experiments, which thus generalizes Chidsey's method for phase transforming particles and porous electrodes. The perfect match between the theoretical Born solvation energy and the fitted reorganization energy validates our assumption in equation (2) that electron transfer is the rate-limiting step in the Faradaic reaction, contrary to the prevailing picture of ion transfer limitation, and also justifies our assumption of including Fermi statistics with constant DOS for the carbon coating electrode. The comparison between MHC and the classic Marcus model reveals the necessity of the Fermi statistics for electrode reactions.

Our electroanalytical method could have broad applicability. We have used micron-sized carbon-coated $LiFePO_4$ to demonstrate the measurement of reaction rates for thin (< 5μm) porous electrodes. For very thick porous electrodes (e.g., >50μm[28]), it becomes difficult to accurately fit the transient current with equation (4), but more sophisticated porous electrode theories[25, 47] can be used to account for transport coupled to phase transformations. Our method can be easily applied to other electrode materials, such as nanoparticles, composites and solid-solution materials. Although the curvature of the Tafel plots for other materials may be different, our method can still extract the rate constant and reorganization energy, which can be compared with microscopic charge-transfer theories to quantitatively understand the reaction mechanism.

Our findings suggest the need for a paradigm shift in the mathematical modeling of kinetics in electrochemical systems. Essentially all models assume BV kinetics, not only for batteries, but also for electrodeposition, corrosion, fuel cells, etc.[47, 48] In spite of the success of electron-transfer theory,[3, 27, 37] the BV equation has been engrained in electrochemistry for over a century, and even used as a "filter to discard models of electron transfer that do not predict [the linear] Tafel's law".[1] Any switch to Marcus or MHC kinetics could have a dramatic effect on model predictions, since the reaction resistance becomes orders of magnitude larger than BV at large overpotentials. It would be particularly interesting to apply generalized Marcus kinetics to non-equilibrium thermodynamics in $LiFePO_4$, which might correct shortcomings of generalized BV kinetics in fitting experimental data at high rates.[49]

## Methods

**Coin cell preparation.** The micron-sized LiFePO$_4$ powder was from A123 System, LLC. Scanning electron microscopy shows that the typical size of the primary particles is around 1μm, and the size of the aggregates is smaller than 5μm. The powder was mixed with carbon black and sodium alginate binder at the weight ratio of 8:1:1, and hand milled to form a smooth slurry. The electrode was prepared by casting the slurry onto an aluminum foil using a doctor blade and dried in a vacuum oven at 100 ℃ overnight. The size of the circular electrodes is 1.27cm in diameter. Loading mass of each electrode is around 1.1mg. The thickness of the electrodes are less than 5μm. Coin cells were assembled in Argon-filled glove boxes (<1 ppm of water and oxygen) with lithium foil as the counter electrode, 1 M LiPF$_6$ in a mixture of ethylene carbonate/diethyl carbonate (EC/DEC, 1:1 by volume) as the electrolyte, and Celgard 3501 as the separator. Galvanostatic charge/discharge performance of the cathode materials was tested in a voltage range of 2.4V-4.2 V. Capacity at C/10 discharge is ~150 mAhg$^{-1}$, calculated on the basis of the mass of LiFePO$_4$ in the composite electrodes.

**Chronoamperometry experiments.** All experiments were conducted with an Arbin battery tester at room temperature. Before each voltage step, coin cells were charged (discharged) at 50μA to 4.0V (2.5V) and then relaxed for 4 hours to reach equilibrium. The size of the voltage step was calculated with respect to the formal potential $E^{0'}$=3.430V, measured as the average of the peak potentials of the cyclic voltammetry. The cut-off current for each step is 5μA.

**Curve Fitting.** Parameters of equation (4) for each transient current were fitted in the Curve Fitting Toolbox of MATLAB. Starting values of $k$, $k_A$, $Q$ and $N_0$ were adjusted to optimize the fitting results under the following principles: (i) the fitted decaying time constant ($1/k$) should be in the same order of magnitude as the time constant estimated directly from the chronoamperogram (time for decaying to 36.8% of the peak current); (ii) largest possible $R^2$ value; (iii) smallest possible 95% confidence bounds for all four parameters; and (iv) the fitted $Q$ should be consistent with the experimental values.

## Acknowledgements

We acknowledge Mr. Tao Gao and Professor Chunsheng Wang at the University of Maryland for providing LiFePO$_4$ coins cells for our experiments, and Dr. Todd R. Ferguson for his inspiring comments on Tafel analysis and Marcus theory.

## Author Contributions

P.B. proposed the method, performed the chronoamperometry experiments, analyzed the data and wrote the manuscript. M.Z.B. guided the theoretical analysis and revised the manuscript.

## Competing financial interests

The authors declare no competing financial interests.

# References

1. Bockris, J.O.M., Reddy, A.K.N., Gamboa-Aldeco, M.E. *Modern Electrochemistry 2A: Fundamentals of Electrodics*, 2nd edn. Kluwer Academic/Plenum Publishers (2000).
2. Bazant, M.Z. Theory of Chemical Kinetics and Charge Transfer based on Nonequilibrium Thermodynamics. *Acc. Chem. Res.* **46**, 1144-1160 (2013).
3. Marcus, R.A. Electron-Transfer Reactions in Chemistry - Theory and Experiment. *Reviews of Modern Physics* **65**, 599-610 (1993).
4. Chidsey, C.E.D. Free-Energy and Temperature-Dependence of Electron-Transfer at the Metal-Electrolyte Interface. *Science* **251**, 919-922 (1991).
5. Padhi, A.K., Nanjundaswamy, K.S., Goodenough, J.B. Phospho-olivines as positive-electrode materials for rechargeable lithium batteries. *J. Electrochem. Soc.* **144**, 1188-1194 (1997).
6. Ravet, N., Chouinard, Y., Magnan, J.F., Besner, S., Gauthier, M., Armand, M. Electroactivity of natural and synthetic triphylite. *J. Power Sources* **97-8**, 503-507 (2001).
7. Zhou, X.F., Wang, F., Zhu, Y.M., Liu, Z.P. Graphene modified LiFePO(4) cathode materials for high power lithium ion batteries. *J. Mater. Chem.* **21**, 3353-3358 (2011).
8. Park, K.S., *et al.* Enhanced Charge-Transfer Kinetics by Anion Surface Modification of LiFePO4. *Chem. Mater.* **24**, 3212-3218 (2012).
9. Maxisch, T., Zhou, F., Ceder, G. Ab initio study of the migration of small polarons in olivine LixFePO4 and their association with lithium ions and vacancies. *Phys. Rev. B* **73**, - (2006).
10. Li, H.Q., Zhou, H.S. Enhancing the performances of Li-ion batteries by carbon-coating: present and future. *Chem. Commun. (Cambridge, U. K.)* **48**, 1201-1217 (2012).
11. Wang, J.J., Sun, X.L. Understanding and recent development of carbon coating on LiFePO4 cathode materials for lithium-ion batteries. *Energy Environ. Sci.* **5**, 5163-5185 (2012).
12. Flandrois, S., Simon, B. Carbon materials for lithium-ion rechargeable batteries. *Carbon* **37**, 165-180 (1999).
13. Srinivasan, V., Newman, J. Discharge model for the lithium iron-phosphate electrode. *J. Electrochem. Soc.* **151**, A1517-A1529 (2004).
14. Bai, P., Cogswell, D.A., Bazant, M.Z. Suppression of Phase Separation in LiFePO4 Nanoparticles During Battery Discharge. *Nano Lett.* **11**, 4890-4896 (2011).
15. Bard, A.J., Faulkner, L.R. *Electrochemical methods : fundamentals and applications*, 2nd edn. John Wiley (2001).
16. Dargaville, S., Farrell, T.W. Predicting Active Material Utilization in LiFePO4 Electrodes Using a Multiscale Mathematical Model. *J. Electrochem. Soc.* **157**, A830-A840 (2010).
17. Safari, M., Delacourt, C. Mathematical Modeling of Lithium Iron Phosphate Electrode: Galvanostatic Charge/Discharge and Path Dependence. *J. Electrochem. Soc.* **158**, A63-A73 (2011).
18. Kasavajjula, U.S., Wang, C.S., Arce, P.E. Discharge model for LiFePO(4) accounting for the solid solution range. *J. Electrochem. Soc.* **155**, A866-A874 (2008).
19. Wang, C.S., Kasavajjula, U.S., Arce, P.E. A discharge model for phase transformation electrodes: Formulation, experimental validation, and analysis. *J. Phys. Chem. C* **111**, 16656-16663 (2007).
20. Marcus, R.A. On Theory of Electron-Transfer Reactions .6. Unified Treatment for Homogeneous and Electrode Reactions. *J. Chem. Phys.* **43**, 679-& (1965).
21. Marcus, R.A. On the Theory of Oxidation-Reduction Reactions Involving Electron Transfer .1. *J. Chem. Phys.* **24**, 966-978 (1956).
22. Munakata, H., Takemura, B., Saito, T., Kanamura, K. Evaluation of real performance of LiFePO4 by using single particle technique. *J. Power Sources* **217**, 444-448 (2012).
23. Kang, B., Ceder, G. Battery materials for ultrafast charging and discharging. *Nature* **458**, 190-193 (2009).
24. Viswanathan, V., Norskov, J.K., Speidel, A., Scheffler, R., Gowda, S., Luntz, A.C. Li-O-2 Kinetic Overpotentials: Tafel Plots from Experiment and First-Principles Theory. *J. Phys. Chem. Lett.* **4**, 556-560 (2013).
25. Ferguson, T.R., Bazant, M.Z. Nonequilibrium Thermodynamics of Porous Electrodes. *J. Electrochem. Soc.* **159**, A1967-A1985 (2012).

26. Dreyer, W., Jamnik, J., Guhlke, C., Huth, R., Moskon, J., Gaberscek, M. The thermodynamic origin of hysteresis in insertion batteries. *Nat. Mater.* **9**, 448-453 (2010).
27. Henstridge, M.C., Laborda, E., Rees, N.V., Compton, R.G. Marcus-Hush-Chidsey theory of electron transfer applied to voltammetry: A review. *Electrochim. Acta* **84**, 12-20 (2012).
28. Bai, P., Tian, G.Y. Statistical kinetics of phase-transforming nanoparticles in LiFePO4 porous electrodes. *Electrochim. Acta* **89**, 644-651 (2013).
29. Cogswell, D.A., Bazant, M.Z. Coherency Strain and the Kinetics of Phase Separation in LiFePO4 Nanoparticles. *ACS Nano* **6**, 2215-2225 (2012).
30. Levi, M.D., *et al.* Collective Phase Transition Dynamics in Microarray Composite LixFePO4 Electrodes Tracked by in Situ Electrochemical Quartz Crystal Admittance. *J. Phys. Chem. C* **117**, 15505-15514 (2013).
31. Laborda, E., Henstridge, M.C., Batchelor-McAuley, C., Compton, R.G. Asymmetric Marcus-Hush theory for voltammetry. *Chem. Soc. Rev.* **42**, 4894-4905 (2013).
32. Oldham, K.B., Myland, J.C. On the evaluation and analysis of the Marcus-Hush-Chidsey integral. *J. Electroanal. Chem.* **655**, 65-72 (2011).
33. Royea, W.J., Hamann, T.W., Brunschwig, B.S., Lewis, N.S. A Comparison between Interfacial Electron-Transfer Rate Constants at Metallic and Graphite Electrodes. *J. Phys. Chem. B* **110**, 19433-19442 (2006).
34. Delmas, C., Maccario, M., Croguennec, L., Le Cras, F., Weill, F. Lithium deintercalation in LiFePO4 nanoparticles via a domino-cascade model. *Nat. Mater.* **7**, 665-671 (2008).
35. Islam, M.S., Driscoll, D.J., Fisher, C.A.J., Slater, P.R. Atomic-scale investigation of defects, dopants, and lithium transport in the LiFePO4 olivine-type battery material. *Chem. Mater.* **17**, 5085-5092 (2005).
36. Wang, L., Zhou, F., Meng, Y.S., Ceder, G. First-principles study of surface properties of LiFePO(4): Surface energy, structure, Wulff shape, and surface redox potential. *Phys. Rev. B* **76**, (2007).
37. Kuznetsov, A.M., Ulstrup, J. *Electron transfer in chemistry and biology : an introduction to the theory*. Wiley (1999).
38. Valoen, L.O., Reimers, J.N. Transport properties of LiPF6-based Li-ion battery electrolytes. *J. Electrochem. Soc.* **152**, A882-A891 (2005).
39. Ouyang, C., Shi, S., Wang, Z., Huang, X., Chen, L. First-principles study of Li ion diffusion in LiFePO_{4}. *Phys. Rev. B* **69**, 104303 (2004).
40. Morgan, D., Van der Ven, A., Ceder, G. Li conductivity in LixMPO4 (M = Mn, Fe, Co, Ni) olivine materials. *Electrochem. Solid-State Lett.* **7**, A30-A32 (2004).
41. Malik, R., Burch, D., Bazant, M., Ceder, G. Particle Size Dependence of the Ionic Diffusivity. *Nano Lett.* **10**, 4123-4127 (2010).
42. Dathar, G.K.P., Sheppard, D., Stevenson, K.J., Henkelman, G. Calculations of Li-Ion Diffusion in Olivine Phosphates. *Chem. Mater.* **23**, 4032-4037 (2011).
43. Zhu, Y.J., Wang, C.S. Galvanostatic Intermittent Titration Technique for Phase-Transformation Electrodes. *J. Phys. Chem. C* **114**, 2830-2841 (2010).
44. Tang, K., Yu, X., Sun, J., Li, H., Huang, X. Kinetic analysis on LiFePO4 thin films by CV, GITT, and EIS. *Electrochim. Acta* **56**, 4869-4875 (2011).
45. Prosini, P.P., Lisi, M., Zane, D., Pasquali, M. Determination of the chemical diffusion coefficient of lithium in LiFePO4. *Solid State Ionics* **148**, 45-51 (2002).
46. Cogswell, D.A., Bazant, M.Z. Theory of Coherent Nucleation in Phase-Separating Nanoparticles. *Nano Lett.* **13**, 3036-3041 (2013).
47. Newman, J.S., Thomas-Alyea, K.E. *Electrochemical systems*, 3rd edn. J. Wiley (2004).
48. Chen, S.L., Liu, Y.W. Electrochemistry at nanometer-sized electrodes. *Phys. Chem. Chem. Phys.* **16**, 635-652 (2014).
49. Dargaville, S., Farrell, T.W. A comparison of mathematical models for phase-change in high-rate LiFePO4 cathodes. *Electrochim. Acta* **111**, 474-490 (2013).